\documentclass[conference, 10pt]{IEEEtran}

\usepackage{fullpage}
\usepackage{indentfirst}
\usepackage{graphicx}
\usepackage{multicol}
\usepackage[ruled,linesnumbered]{algorithm2e}
\usepackage{comment}
\usepackage{amsmath}
\usepackage{array}
\usepackage{epstopdf}
\usepackage{rotating}
\usepackage{booktabs}
\usepackage{stmaryrd}
\usepackage{tabularx}
\usepackage{pdflscape}
\usepackage{amssymb}
\usepackage[table,xcdraw]{xcolor}
\usepackage{tabularx}

\begin{document}

\title{An Integrated Recommender Algorithm for Rating Prediction}
\author{Yefeng Ruan \\yefruan@iupui.edu \and Tzu-Chun (Vanson) Lin \\lintz@iupui.edu.tw} 
 
\date{}
\maketitle

\begin{abstract}
Recommender system is currently widely used in many e-commerce systems, such as Amazon, eBay, and so on. It aims to help users to find items which they may be interested in. In literature, neighborhood-based collaborative filtering and matrix factorization are two common methods used in recommender systems. In this paper, we combine these two methods with personalized weights on them. Rather than using fixed weights for these two methods, we assume each user has her/his own preference over them. Our results shows that our algorithm outperforms neighborhood-based collaborative filtering algorithm, matrix factorization algorithm and their combination with fixed weights.  
\end{abstract}

\section{Introduction}
Recommender systems are now widely deployed in may e-commerces, like Amazon, eBay, Epinions and so on, as these platforms become more and more popular. The main purpose of recommender systems is to provide users a list of items that they may be interested in. Items in the list are ranked based on metrics like similarity, relevance and so on. Such type of recommender systems are often mentioned as top-N recommendations \cite{Jamali:2009:UTN:1639714.1639745} \cite{Cremonesi:2010:PRA:1864708.1864721} \cite{Zhao:2014:LSC:2661829.2661998}. Besides ranked list, in some case, researchers are also interested in predicting rating that users will rate for items. This is always called rating prediction problem, and many works belong to this category \cite{Massa2007} \cite{Liu:2013:SSN:2488388.2488457} \cite{Ma:2008:SSR:1458082.1458205} \cite{Nazemian:2012:IMT:2456719.2456998} \cite{Rendle:2012:LRS:2124295.2124313} \cite{Ma:2014:IRA:2661829.2662085}. 

Recommendation approaches can be basically divided into two categories: content based and collaborative filtering based approaches \cite{yang2014survey}. Among them collaborative filtering based approach are wildly used by many works \cite{Massa2007} \cite{Koren:2010:CFT:1721654.1721677} \cite{Ma:2014:IRA:2661829.2662085} \cite{Ma:2008:SSR:1458082.1458205}. Collaborative filtering based approach can be further divided into neighborhood-based (or memory based) and model based collaborative filtering approach \cite{yang2014survey}. Classical neighborhood-based collaborative filtering approaches assume similar users (neighbors) have similar tastes on items such that their purchase or rating behaviors are also very similar \cite{breese1998empirical}. In traditional neighborhood-based collaborative filtering approach, user-user (item-item) similarities are calculated by two users' previous purchase or rating behaviors (two items' common buyers). While model based collaborative filtering approach, like Matrix Factorization \cite{10.1109/MC.2009.263}, model both users and items with some latent factors, and these latent factors are learned in the training stage. In this paper, we propose a new algorithm which integrates neighborhood-based collaborative filtering (CF) and Matrix Factorization (MF). When considering these two methods together, rather than assign them with fixed weights for all the users, we assume that each user has her/his own preference over them.

In this paper, the Twitter-based movie rating dataset, MovieTweeting \cite{dooms2013movietweetings}, is chosen for the recommender system development. The rating data are extracted from Twitter, in which users rated a movie on IMDB, and posted the score on their Twitter timeline. The organizer of MovieTweeting extracts the ratings as well as relevant information from Twitter, and posts the dataset on Internet, inviting public to develop customized recommender system based on this dataset. 

The interesting part of the dataset is the access to real social behavior for each user, led by the real Twitter IDs provided in the dataset. Meanwhile, the dataset also gives the entries of the rated movie, by IMDB movie IDs. Unlike usual recommender dataset, which only contains rating values associated to anonymous users and items, the MovieTweeting dataset gives linkages to the real world, enable the developers to search more probabilities in potential useful data.

In the following of this paper, we first introduce the background and related works in Section \ref{sec2}. We analyze the social connections among users in Section\ref{sec3}. After investigation, the proposed social network is too sparse. Also we review two very classical algorithms in recommender systems: neighborhood-based collaborative filtering and matrix factorization in Section \ref{sec4}. In Section \ref{sec5}, we integrated these two methods together with another baseline method, and incorporate them into a single model. Furthermore, we proposed a improved version, in which we assume that each user has her/his own preference over these two methods. We call it integrated algorithm 2.0. In Section \ref{sec6}, experimental results show that integrated algorithm can perform much better than neighborhood-based collaborative filtering and matrix factorization individually. Also by taking users' different preferences over these methods into account, we can even achieve better. We conclude this paper in Section \ref{sec7}

\section{Background and Related Works}\label{sec2}
\subsection{Collaborative Filtering}
Collaborative filtering is the most wildly used approach and has acceptable accuracy in many cases \cite{yang2014survey}. And it can be divided into neighborhood-based and model based approach. In this paper we mainly focus on and use neighborhood-based approach. To recommend items to a user, it assumes that similar users have similar tastes such that items purchased by similar users may also be of interest for her. Similarly, based on user's previous purchased items, similar items might also be attractive to her. Based on which type of similarity measured, it can further be divided into user base and item based approach. Collaborative filtering tries to find out similar users or items using users' previous purchase behaviors. Therefore, it takes user-item rating or purchase matrix ($M$) as input, in which each row represents a user and each column represents an item. For example, in the input matrix, $M_{ij}$ represents user $i$'s rating or purchase behavior on item $j$.

Besides neighborhood-based approach, matrix factorization is another popular approach used in recommender systems. It assumes that users and items have the same amount of hidden features. Therefore, both users and items can be represented by $m * n$ matrix, in which m is the number of users/items, and n is the number of hidden features. Then, the prediction becomes the matrix factorization problem.

\subsection{Related Works}
As seen by many researchers, traditional collaborative filtering approach have the data sparsity issue and does not solve cold starts problem very well, there are many works proposed to solve this problem. 

TaRS \cite{Massa2007} uses collaborative filtering approach along with social trust information to produce advice. It uses trust propagation -- MoleTrust \cite{massa2007trust} to infer indirect trust among users such that more users can be connected and increase the coverage. Based on TaRS, \cite{Nazemian:2012:IMT:2456719.2456998} proposed a model which also takes distrust into account as well. \cite{breese1998empirical} also use trust metric as weights, but at the same time it keeps similarity. It filters out links in the trust network if two users' similarity is below a threshold. 

Using similarity and social trust information, \cite{Ma:2014:IRA:2661829.2662085} can even cluster users into groups and find groups' behavior patterns instead of single user's behavior pattern. \cite{yuan2010improved} learned social networks' small world property and also cluster users together in order to do better prediction.

The most similar work with our new algorithm is \cite{Koren:2008:FMN:1401890.1401944}. It Combines neighborhood-based and model based approach, which can use either explicit or implicit social information in recommender systems. 

\section{Social Trust Network}\label{sec3}
The decisions people make are usually influenced by others, especially the ones they trust. This idea constructs the so-called Social Trust Network \cite{Massa2007}. It was difficult to construct such network back couple decades ago since most information was not quantified, while such network construction becomes available nowadays since the trust between people can be observed from their behavior on social network website.

Given real Twitter IDs from the dataset provided, we are able to connect the movie ratings to the raters' real life. On Twitter, a well-known social network website, user can retweet other user's tweet. The retweeting behavior makes trust between people observable. We can make a guess, if user A retweets user B's tweet, we can assume user A tends to trust user B. This could help us predict a possible movie rating that is not rated by user A but is rated by user B who is trusted by user A.

Thus, a trust indicator between a pair of user can be formulated as Equation \ref{eq::trustiness},

\begin{equation}
\label{eq::trustiness}
Trustiness(i,\ j) = \frac{n(i,\ j)}{C}
\end{equation}

while, the $n$ represents the number of retweets posted by user i from user j. The more posts user i has retweeted user j, the high level trust of user i toward user j. If the trust network is observed valid, the resulting trust factor can later be integrated to overall mathematical model.

\subsubsection{Attempt of Network Construction}
We first extracted social content posted on the Twitter of each user in the training set. A clip of the retweeting data is shown in Table \ref{tab:retweetdata}.

\begin{table*}[htbp]
\centering
\caption{A Clip of Retweeting Data}
\label{tab:retweetdata}
\begin{tabular}{@{}lllll@{}}
\toprule
Username         & Location      & Retweeted Username & Retweet Content                                                               & Timestamp                                                              \\ \midrule
GreatBritain\_GB & Great Britain & sonianitiwadee     & \begin{tabular}[c]{@{}l@{}}Cooler than he :\\ Adidas...\end{tabular}          & \begin{tabular}[c]{@{}l@{}}Tue Oct 15 00:49:16\\ EDT 2014\end{tabular} \\
kvakke           & Oslo, Norway  & netliferesearch    & i dag. Veldig l�rerikt. :)                                                    & \begin{tabular}[c]{@{}l@{}}Mon Nov 03\\ 05:28:35 EST 2014\end{tabular} \\
luisferreras     & Dime          & FernandoSued1      & \begin{tabular}[c]{@{}l@{}}Dios mio, asi quedo el\\ veh de Oscar\end{tabular} & \begin{tabular}[c]{@{}l@{}}Sat Oct 25 21:44:40\\ EDT 2014\end{tabular} \\ \bottomrule
\end{tabular}
\end{table*}

The first column contains the users provided in the MovieTweeting dataset, while the third column contains the users who are retweeted by the users in first column. We organize the data and input to Gephi, an open-source software, to plot network diagram using Fruchterman Reingold Algorithm to observe the behavior of that network, as Figure \ref{fig:network} below.

\begin{figure}[!htbp]
\centering
\includegraphics[width=0.48\textwidth, keepaspectratio]{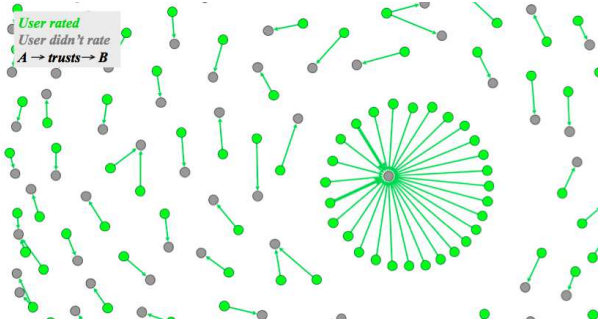}
\caption{A Glimpse of the Trust Network}
\label{fig:network}
\end{figure}

Note that each user we focus for Twitter content extraction is from MovieTweeting dataset. All the users in the first column did provide rating information (marked green). However, the users in the third column, or the people being retweeted (marked grey), do not necessarily exist in the MovieTweeting dataset. In other word, the people being trusted in the network did not rate any movie for grant. However, we should only care those being retweeted as well as providing rating information, since only those being trusted can help us predict those followers' movie preferences. 

After investigation, only 0.1\% of those being trusted did provide movie rating (also exist in the training dataset). And only 0.5\% of training set user are influenced by trust network. The sparse network impact is way too small. Thus, this approach is discarded from the overall model.

\section{Collaborative Filtering}\label{sec4}
\subsection{Neighborhood-based Collaborative Filtering}
Neighborhood-based collaborative filtering (CF) \cite{breese1998empirical} is one of the most classical recommendation system algorithm. Basically it assumes that similar users have similar tastes on items, or similar items will attract same users. To measure similarity among users or items, we use cosine similarity as shown in Equations \ref{eq::user_sim} and \ref{eq::item_sim}. 

In calculating cosine similarity, we deduct user or item's average rating $\bar{r_u}$ or $\bar{r_i}$ from $r$. This is because deviations from average ratings are more useful in inferring users' preferences. Ratings can not directly reflect users' preference, as it will be affected by users' basic favor (average ratings). For example, if user $A$ rates item $i$ for 4, and his average rating is also 4. For user $B$, he rates item $i$ for 3, but his average rating is 2. In this example, user $B$ shows more favor in item $i$ than user $A$. Therefore, it is necessary to remove average ratings from $r$ when considering user or item similarity.

Besides this, we also take the number of common items rated by two users (or the number of common users rating two items) into account. It is obvious that given the same deviation, more common items two users rate, more similar they are. So we introduce the second term in Equations \ref{eq::user_sim} and \ref{eq::item_sim}. $C_{uv}$ is the set of items which rated by both user $u$ and $v$, and $C_{ij}$ is the set of users who rate both item $i$ and $j$.  

\begin{equation}
\small
\label{eq::user_sim}
\begin{split}
s(u,v) =& \frac{\sum_{i \in C_{uv}} (r(u,i) - \bar{r_i}) * (r(v,i) - \bar{r_i})}{\sqrt{\sum_{i \in C_{uv}}(r(u,i) - \bar{r_i})^2} * \sqrt{\sum_{i \in C_{uv}} (r(v,i) - \bar{r_i})^2}} \\
&* \frac{|C_{uv}|}{|C_{uv}| + 100}
\end{split}
\end{equation}

\begin{equation}
\small
\label{eq::item_sim}
\begin{split}
s(i,j) =& \frac{\sum_{u \in C_{ij}} (r(u,i) - \bar{r_u}) * (r(u,j) - \bar{r_u})}{\sqrt{\sum_{u \in C_{ij}}(r(u,i) - \bar{r_u})^2} * \sqrt{\sum_{u \in C_{ij}} (r(u,j) - \bar{r_u})^2}} \\
&* \frac{|C_{ij}|}{|C_{ij}| + 100}
\end{split}
\end{equation}

Intuitively, more similar user or item is, more important the corresponding rating is. Based on what similarity metrics used, CF can be divided into two categories: user-based and item based. Equations \ref{eq::user_pre} and \ref{eq::item_pre} show prediction function of user-based and item-based separately. Here $b(u,i)$ is the baseline from average ratings, as shown in Equation \ref{eq::baseline}.

\begin{equation}
\small
\label{eq::user_pre}
\widehat{r(u,i)} = b(u,i) + \frac{\sum_{v \in s_u^N\bigcap r(v,i)\neq\emptyset} s(u,v)*(r(v,i) - b(v,i))}{\sum_{v \in s_u^N\bigcap r(v,i)\neq\emptyset} s(u,v)}
\end{equation}

\begin{equation}
\small
\label{eq::item_pre}
\widehat{r(u,i)} = b(u,i) + \frac{\sum_{j \in s_i^N\bigcap r(u,j)\neq\emptyset} s(i,j)*(r(u,j) - b(u,j))}{\sum_{j \in s_i^N\bigcap r(u,j)\neq\emptyset} s(i,j)}
\end{equation}

\begin{equation}
\label{eq::baseline}
b(u,i) = \bar{r} + \bar{r_u} + \bar{r_i}
\end{equation}

In real applications, we only consider top $N$ most similar users or items when we do prediction. We will see how $N$ affects CF's performance on out dataset.

\subsection{Matrix Factorization}
Matrix factorization (MF) \cite{billsus1998learning} is another popular algorithm for recommender systems. Unlike neighborhood-based collaborative filtering, it does not require semantic explanation and no domain expert needed. Although it also assumes that there exists user preference factors and item feature factors, they are not directly available. Instead, it assume there are certain number of hidden factors, which capture users' features and items' features. 

In MF, we assume there are $K$ hidden factors to model users' preference. Then users can be represented by a $M*K$ matrix, which is called $P$. Here $M$ is the number of users. Each row of $P$ represents a user and each column of $P$ represents one users' feature. In order to relate users with items, items are also represented by a matrix $Q$, which has $N$ rows and also $K$ columns. In such way, predicted rating can be written as Equation \ref{eq::MF_pre}.

\begin{equation}
\label{eq::MF_pre}
\widehat{r(u,i)} = \bar{r} + p_u q_i^T
\end{equation}

We show objective function in Equation \ref{eq::MF_obj}. It includes two parts: error term and regularization term. Regularization term is used to avoid over fitting. To solve this problem, we use alternating least squares for $P$ and $Q$. For each rating pair $(u,i)$ in the training dataset, we update $P(u)$ and $Q(i)$ according to Equations \ref{eq::MF_P} and \ref{eq::MF_Q}.

\begin{equation}
\label{eq::MF_obj}
f(P,Q) = \sum_{(u,i) \in R} (r(u,i) - \widehat{r(u,i)})^2 + \lambda * (||P||_F^2 + ||Q||_F^2)
\end{equation}

\begin{equation}
\label{eq::MF_P}
P(u) = [\sum_{(u,i) \in R} Q(i) r(u,i)] * [\lambda I +\sum_{(u,i) \in R} Q(i)^TQ(i)]^{-1}
\end{equation}

\begin{equation}
\label{eq::MF_Q}
Q(i) = [\sum_{(u,i) \in R} P(u) r(u,i)] * [\lambda I +\sum_{(u,i) \in R} P(u)^TP(u)]^{-1}
\end{equation}

To get the optimal solution, we iterate $R$ again and again, until it converges. Note that in each iteration all rating pairs $(u,i)$ are visited. After each iteration ends, we compare objective function with the previous iteration, if it does not change a lot, we consider it is converged. 
\[
 f =\begin{cases}
     converged, \;\;\;\;\;\;\;\;\;\;\;\; If \frac{|f(t)-f(t-1)|}{f(t-1)} \leq \epsilon \\
    not \;\;converged, \;\;\;\;\;\; Otherwise 
\end{cases}
\]

For all iterations, we print out their $MAE$s and we select the minimum one as our result. Also we know that the number of latent factors $K$ can affect both $MAE$ and time complexity, we will see how can $K$ affect results later.

\section{Integrated Algorithm}\label{sec5}
\subsection{Integrated CF and MF 1.0}
neighborhood-based collaborative filtering and matrix factorization are two widely used algorithms in recommender systems. However, both of them themselves are not perfect. When calculating user or item similarities in neighborhood-based collaborative filtering algorithm, it only consider two users' or two items' common ratings, all other rating information is not used at all. On the other hand, matrix factorization leverages all users' and items' ratings to model their features. But it does not take user-user or item-item relationship into account. To overcome this problem, like in in \cite{Koren:2008:FMN:1401890.1401944} we integrate these two algorithms together in a single model. The prediction functions in Equation \ref{eq::CF_MF1_pre} contains three terms: bias baseline, matrix factorization predicted deviation from bias, and neighborhood-based collaborative filtering predicted deviation from bias.

\begin{equation}
\label{eq::CF_MF1_pre}
\begin{split}
{} &\widehat{r(u,i)} = \bar{r} + bu(u) + bi(i) + p_u q_i^T + \\
&\sum_{j \in s_i^N\bigcap r(u,j)\neq\emptyset} (r(u,j) - b(u,j))*w(i,j) 
\end{split}
\end{equation}

To incorporate these two algorithms, it is intuitive for matrix factorization, but we modify a little bit for the neighborhood-based collaborative filtering part. In the traditional neighborhood-based collaborative filtering algorithm, weights $w(i,j)$ is user dependent. How much item $j$ will affect user $u$'s rating on item $i$ does not only depend on $s(i,j)$, but also $i$'s similarities with other items. But as stated in \cite{Bell2007}, it is helpful to make these weights global and user independent. In such a way, $w(i,j)$ is treated as variables, and we can learn it in the training stage. Here $j$ is among the top-N similar items (or users) of item $i$, for each $(u,i)$ rating pair in the training dataset. Selection is based on the similarity metric mentioned in Equation \ref{eq::item_sim} for pairs of items. Also we assume $bu$ and $bi$ are variables, which means users' and items' bias are changing over time. Therefore, given the prediction function, we can write our objective function as in Equation \ref{eq::CF_MF1_obj}. 

\begin{equation}
\label{eq::CF_MF1_obj}
\begin{split}
{} &f(bu,bi,P,Q,w) = \sum_{(u,i) \in R} (r(u,i) - \widehat{r(u,i)})^2 +\lambda_1 *\\ &(||bu||^2 + ||bi||^2) + \lambda_2 * (||P||^2 + ||Q||^2) +\lambda_3 * ||w||^2  
\end{split}
\end{equation}

Again $\lambda_1$, $\lambda_2$ and $\lambda_3$ are regularization parameters. Therefore, our goal is to minimize the objective function. To solve this optimization problem, we use Stochastic gradient descent method. Instead of calculating gradients over whole training dataset, we approximate it at single examples. And we will update variables for each given training pairs. The learning rate is controlled by parameters $r1$, $r2$ and $r3$. If we denote error between predicted rating and actual rating as $eui$ for user and item pair $(u,i)$ in the training dataset, updating process can be written as following.

\begin{gather*}
\label{eq::CF_MF1_update}
\begin{split}
bu(u) &= bu(u) - r1* \frac{\partial f}{\partial bu(u)} = {}\\
& bu(u) - r1*(\lambda_1*bu(u) - eui) 
\end{split} \\
\begin{split}
bi(i) &= bi(i) - r1* \frac{\partial f}{\partial bi(i)} = {}\\
& bi(i) - r1*(\lambda_1*bi(i) - eui)  
\end{split}
\\
p_u = p_u - r2* \frac{\partial f}{\partial p_u} = p_u - r2*(\lambda_2*p_u - eui*q_i)  \\
q_i = q_i - r2* \frac{\partial f}{\partial q_i} = q_i - r2*(\lambda_2*q_i - eui*p_u)  \\
\begin{split}
w(i,j) = &w(i,j) - r3* \frac{\partial f}{\partial w(i,j)}= w(i,j) - {}\\
& r3*(\lambda_3*w(i,j) - eui*(r(u,j)-b(u,j)))
\end{split}
\end{gather*}

We continue to update these variables until it converges, which means the objective function remains stable.

\subsection{Integrated CF and MF 2.0}
In the above integrated algorithm, we assume bias baseline, neighborhood-based collaborative filtering and matrix factorization are equally important such that we just simply add them together. But it can be the case that different users may favor different methods among these three algorithms. For example, user $A$'s behaviors may be very similar with bias baseline such that neighborhood-based collaborative filtering and matrix factorization should not affect much. However, for user $B$, it is possible that his rating behavior is more similar with matrix factorization than other two. We realize that it is necessary to model users' preferences over three methods. Therefore we put user-based weights ($a1$, $a2$ and $a3$) for three methods. The prediction function is in Equation \ref{eq::CF_MF2_pre}. 

\begin{equation}
\label{eq::CF_MF2_pre}
\begin{split}
\widehat{r(u,i)} &= a1(u) * (\bar{r} + bu(u) + bi(i)) + a2(u) * p_u q_i^T +{}\\
& a3(u) * \sum_{j \in s_i^N\bigcap r(u,j)\neq\emptyset} (r(u,j) - b(u,j))*w(i,j) 
\end{split}
\end{equation}

Correspondingly, its objective function can be written as \ref{eq::CF_MF2_obj}.

\begin{equation}
\label{eq::CF_MF2_obj}
\begin{split}
{}&f(bu,bi,P,Q,w) = \sum_{(u,i) \in R} (r(u,i) - \widehat{r(u,i)})^2 + {}\\
&\lambda_1 * (||bu||^2 + ||bi||^2)+ \lambda_2 * (||P||^2 + ||Q||^2) +{}\\
&\lambda_3 * ||w||^2 + \lambda_4 * (||a1||^2 + ||a2||^2 + ||a3||^2)
\end{split}
\end{equation}

Similarly, we use Stochastic gradient descent method to solve this optimization problem, updating process can be seen as following. And we show this method in Algorithm \ref{alg::CF_MF2}.
\begin{gather*}
bu(u) = bu(u) - r1* \frac{\partial f}{\partial bu(u)} \\
bi(i) = bi(i) - r1* \frac{\partial f}{\partial bi(i)} \\ 
p_u = p_u - r2* \frac{\partial f}{\partial p_u} \\
q_i = q_i - r2* \frac{\partial f}{\partial q_i} \\
w(i,j) = w(i,j) - r3* \frac{\partial f}{\partial w(i,j)} \\
a1(u) = a1(u) - r4* \frac{\partial f}{\partial a1(u)} \\
a2(u) = a2(u) - r4* \frac{\partial f}{\partial a2(u)} \\
a3(u) = a3(u) - r4* \frac{\partial f}{\partial a3(u)} \\
\end{gather*}

{\fontsize{10}{7}\selectfont
\begin{algorithm}
\label{alg::CF_MF2}
\SetKwInput{KwData}{Input}
\SetKwInput{KwResult}{Output}
 \KwData{K: latent dimension, N: top N similar items, $\lambda$: constant parameter (100) in similarity calculation, maxIter: maximum number of iterations, $\epsilon$: converge condition, $\lambda1,\lambda2,\lambda3,\lambda4$: regularization parameters, $r1,r2,r3,r4$: learning rates, R: training data, T: testing data}
 \KwResult{MAE}
 Calculate average ratings $\bar{r}$, $bu$, $bi$ \;
 Initialize P and Q with $\frac{1}{K}$\;
 Initialize $a1$, $a2$ and $a3$ with $1$\;
 \For{$i = 1$ \textbf{to} $Item\_Size$}{
 	\For{$j=i+1$ \textbf{to} $Item\_Size$}{
 		Calculate $s(i,j)$ \;
  	}
 }
 \For{$i = 1$ \textbf{to} $Item\_Size$}{
 	Sort and select top N similar items $s_i^N$\;
 	Initialize $w$ with similarity score\;
 }
 t=0\;
 \While{$t<maxIter$}{
 	++t\;
	\For{all $(u,i) \in R$ }{
 		$e_{ui} = r(u,i) - \widehat{r(u,i)}$\;
		$R_{u,i}^N = R_u \bigcap s_i^N$\;
		$bu(u) = bu(u) - r1*(\lambda1*bu(u) - e_{ui})$\; 
		$bi(i) = bi(i) - r1*(\lambda1*bi(i) - e_{ui})$\;
		$p_u = p_u - r2*(\lambda2*p_u - e_{ui}*q_i)$\;
		$q_i = q_i - r2*(\lambda2*q_i - e_{ui}*p_u)$\;
		\For{each $j \in R_{u,i}^N$}{ 
 			$w(i,j) = w(i,j) - r3*(\lambda3*w(i,j)-e_{ui}*(r(u,j) - b(u,j)))$\;
 		}
 		$a1(u) = a1(u) - r4*(\lambda4*a1(u) - e_{ui}*(\bar{r} + bu(u) + bi(i)))$\;
 		$a2(u) = a2(u) - r4*(\lambda4*a2(u) - e_{ui}*(p_u q_i^T))$\;
 		$a3(u) = a3(u) - r4*(\lambda4*a3(u) - e_{ui}*(\sum_j (r(u,j)-b(u,j))*w(i,j)))$\;
 		
 	}
 	Calculate MAE on T\;
	\If{$\frac{|f(t)-f(t-1)|}{f(t-1)} < \epsilon$}{
 		break\;
 	}
 }
 return MAE\;
 \caption{Integrated CF and MF algorithm}
\end{algorithm}
}

%%%%%%%%%%%%%%%% 
% Result and conclusions
%%%%%%%%%%%%%%%%
\section{Results and Conclusions}\label{sec6}
\subsection{Dataset Description}
We use the same dataset -- MoiveTweeting as \cite{dooms2013movietweetings} \cite{Said:2014:RSC:2645710.2645779}. It is a up-to-date dataset and we download it on Nov 7, 2014. It is like that we take a snapshot of the dataset at that time. MoiveTweeting collects all tweets from Twitter having the format ``*I rated \#IMDB*''. In such tweets, it extracts user ids and movie ids, associated with ratings. Therefore, it can be seen as a user-item purchase matrix. It is also the dataset used in the ACM RecSys Challenge 2014.

Originally, the dataset contains 22,079 users and 13,618 items in the training dataset. But we find that some of them do not appear in the testing dataset. On the other hand some users and items in the testing dataset never appear in the training dataset. Therefore, we remove such kind of users and items from the dataset. After pruning, details of the dataset can be seen in Table \ref{tab:data}.

\begin{table}[h!]
\caption{Dataset details}
\label{tab:data}
\centering
\begin{tabular}{cc}
\hline
\# of total users&24,924 \\
\# of total items& 15,142 \\
\# of users in R& 22,079\\
\# of items in R&13,618\\
\# of pairs in R&170,285\\
\# of users in T&5,717\\
\# of items in T&4,226 \\
\# of pairs in T&16,848 \\
\hline
\end{tabular}
\end{table} 

\subsection{Performance Comparison}
In this section, we compare our integrated algorithms with neighborhood-based collaborative filtering and matrix factorization methods. 

\subsubsection{Parameters}
For the MF method, we let $\lambda$ in Equation \ref{eq::MF_obj} equals to 10, as it gives us the best performance. And for algorithms using stochastic gradient descent method, we set $\epsilon = 0.0001$ and $maxIter = 100$. In CF and MF integrated algorithm 1.0, we set $\lambda1 = 0.1$, $\lambda2 = 0.1$, $\lambda3 = 1.0$, $r1 = 0.002$, $r2 = 0.005$, $r3 = 0.002$. And for integrated algotrithm 2.0, we set $\lambda1 = 0.1$, $\lambda2 = 0.1$, $\lambda3 = 1.0$, $\lambda4 = 1.0$, $r1 = 0.002$, $r2 = 0.01$, $r3 = 0.002$, $r4 = 0.002$.

\subsubsection{Comparisons}
\begin{figure}[!htbp]
\centering
\includegraphics[width=0.48\textwidth, keepaspectratio]{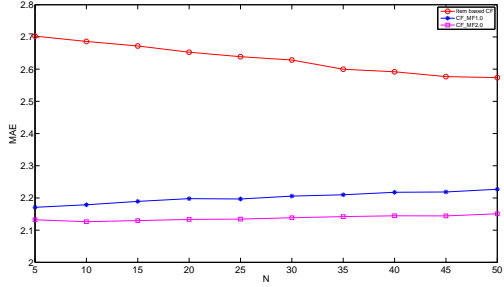}
\caption{Comparison of CF, CF\_MF1.0 and CF\_MF2.0 integrated algorithms}
\label{fig:performance1}
\end{figure}

\begin{figure}[!htbp]
\centering
\includegraphics[width=0.48\textwidth, keepaspectratio]{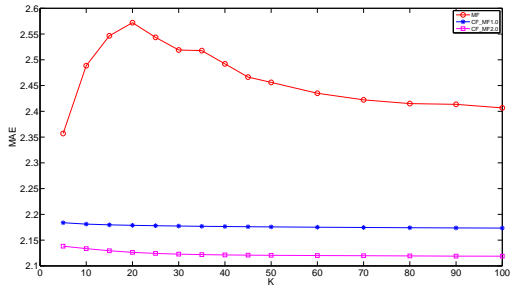}
\caption{Comparison of MF, CF\_MF1.0 and CF\_MF2.0 integrated algorithms}
\label{fig:performance2}
\end{figure}

We compare the integrated methods with neighborhood-based collaborative filtering and matrix factorization methods separately. In these two comparisons, we increase N and K. We compare CF, CF\_MF1.0 and CF\_MF2.0 algorithms' performances when we increase N from 5 to 50, with K fixed at 20 for CF\_MF1.0 and CF\_MF 2.0 algorithms. Similarly, we increase K from 5 to 100 to compare MF, CF\_MF1.0 and CF\_MF2.0 algorithms' performances with N fixed at 10. Prediction error MAE are shown in Figure \ref{fig:performance1} and \ref{fig:performance2} separately.

From Figure ref{fig:performance1} we can see that our integrated methods achieve more than $15\%$ improvement over neighborhood-based collaborative filtering. we also note that when we increase N, the results do not change a lot. We can see from Figure \ref{fig:performance2} that CF\_MF1.0 algorithm improve accuracy by more than $10\%$ than traditional matrix factorization method. And CF\_MF2.0 can even achieve more than CF\_MF1.0. This means that by assuming that users have different favors over three methods can achieve more improvement than equally treating them. 

\begin{figure}[!htbp]
\centering
\includegraphics[width=0.48\textwidth, keepaspectratio]{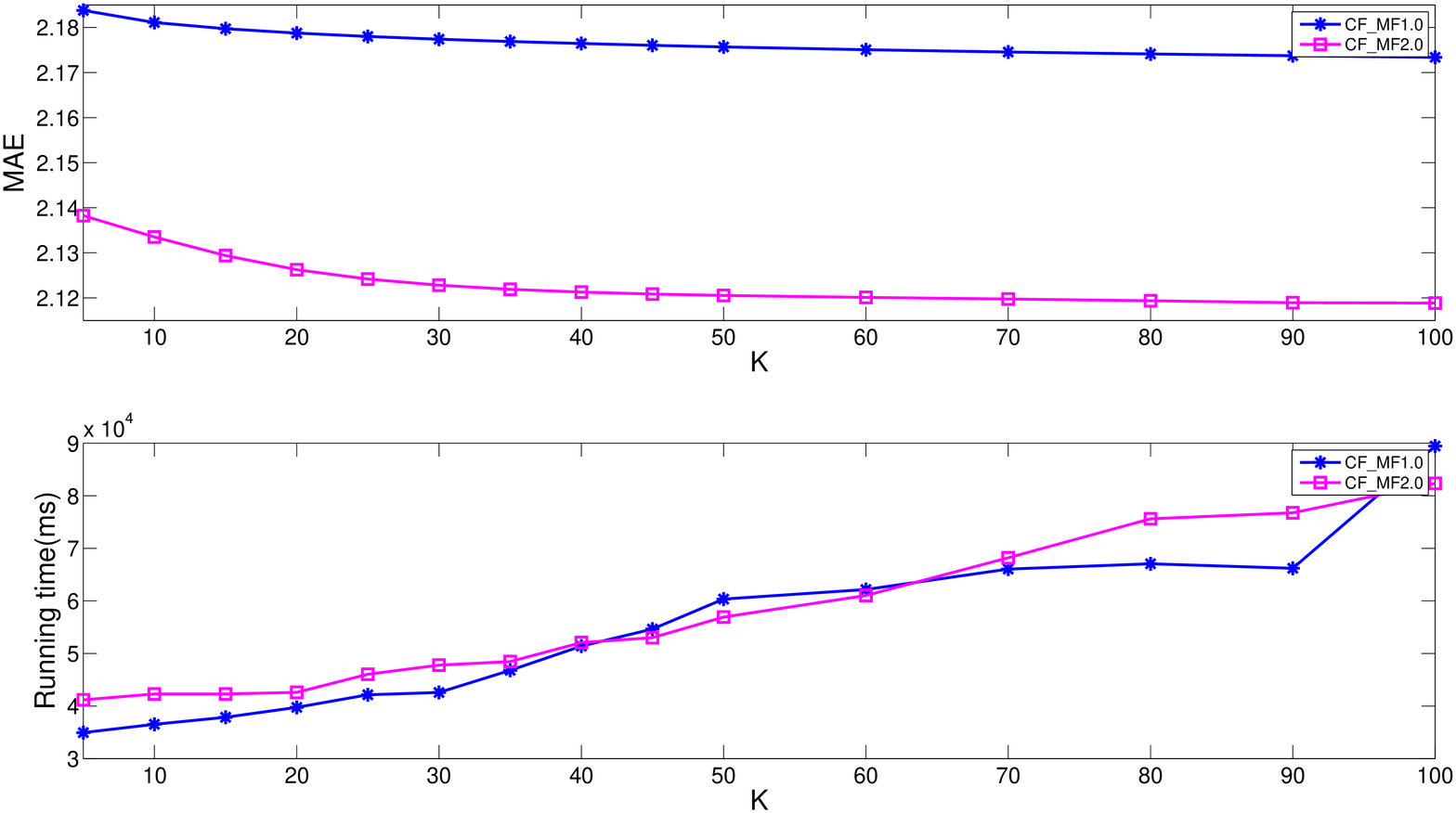}
\caption{CF\_MF1.0 and CF\_MF2.0 MAE and running time}
\label{fig:performance3}
\end{figure}

In order to illustrate more clearly K's influence on algorithms' performance, we show CF\_MF1.0 and CF\_MF2.0's performance again in Figure \ref{fig:performance3}, along with their running time. We can see that increasing K can help us to reduce error. However, at the same time running time is also increasing. Therefore there exists trade-off between prediction accuracy and running time. We list MAEs of CF\_MF1.0 and CF\_MF2.0 in Table \ref{tab::per}.  

\begin{table}[h!]
\caption{CF\_MF1.0 and CF\_MF2.0's MAE}
\label{tab::per}
\centering
\begin{tabular}{ccc}
\hline
K&CF\_MF1.0 & CF\_MF2.0  \\
\hline
5&2.18379&2.13825 \\	
10&2.18109&2.13351\\
15&2.17971&2.12938\\
20&2.17875&2.12625\\	
25&2.17801&2.12418\\
30&2.17739&2.12281\\		
35&2.17688&2.12189\\
40&2.17643&2.12127\\		
45&2.17604&2.12084\\
50&2.17568&2.12053\\		
60&2.17507&2.12010\\
70&2.17455&2.11974\\
80&2.17411&2.11937\\
90&2.17372&2.11893\\	
100&2.17337&2.11884\\
\hline
\end{tabular}
\end{table} 

\section{Conclusions and Future Works}\label{sec7}
In this paper, we propose a new algorithm which integrates neighborhood-based collaborative filtering (CF) and Matrix Factorization (MF). When considering these two methods together, rather than assign them with fixed weights for all the users, we assume that each user has her/his own preference over them. Our results on the MovieTweetings dataset shows that our algorithm outperforms neighborhood-based collaborative filtering algorithm, matrix factorization algorithm and their combination with fixed weights.

For integrated algorithms, we can still do parameter evaluations based on evaluation dataset. Also we may consider some constrains on variables, like $a1$, $a2$, $a3$ and $w$. The integrated algorithms are flexible, it is easy to add other terms, like social side information into it. 

The future work will be focused on the social network analysis. Since the relations between pairs of users do not work due to sparsity, attention should be paid to individual background information, such as age, location, gender, education level, etc.

\section{Work Distribution}
This work is based on a course project (Recommender systems, IUPUI, Fall, 2014).

Yefeng Ruan extracts users' tweets from Twitter, implements CF and MF algorithms separately, proposes and implements CF and MF integrated algorithms 1.0 and 2.0, also compares and analyzes results.

Tzu-Chun Lin analyzes the social relationship among users and also implement SVD algorithm. But as it is same as MF algorithm, we do not present it here.
\bibliographystyle{ieeetr}
\bibliography{final_report}
\end{document}